\documentclass[a4paper, 10pt]{article}
\pdfoutput=1

\usepackage{amsmath, amssymb, bm}
\usepackage[top=1in, bottom=1.2in, left=0.5in, right=0.5in]{geometry}
\usepackage[]{graphicx}
\usepackage[font=scriptsize,labelfont=bf]{caption}
\usepackage[caption=false]{subfig}
\usepackage{textcomp}
\usepackage{authblk}
\usepackage{hyperref}

\newcommand{\bs}{\boldsymbol}

\begin{document}

\title{\Large{\bf{Graviton-photon mixing. \\ Exact solution in a constant magnetic field}}}
\author{Damian Ejlli}

\affil{\emph{\normalsize{School of Physics and Astronomy, Cardiff University, The Parade, Cardiff CF24 3AA, United Kingdom}}}

\date{}

\maketitle

\begin{abstract}
In this work I study the effect of graviton-photon mixing in a constant external magnetic field and find for the first time in the literature exact solution of the equations of motion. In particular, I apply the effect of graviton-photon mixing to the case of interaction of gravitational waves with an external magnetic field and calculate the intensity Stokes parameter of the produced electromagnetic radiation. The obtained  results are new and extend previous results obtained by using approximation methods. 
\end{abstract}

\section{Introduction}
\label{sec:1}

In the recent years there have been several detections of Gravitational Waves (GWs) by LIGO and/or VIRGO from astrophysical sources such as binary systems of black holes and neutron stars. The detection of GWs has been one of the biggest experimental achievements of the century. The GWs detected by LIGO and/or VIRGO typically have a frequency range from few Hz up to few kHz. In addition, there are many other sources that emit GWs in other frequency bands with equally physical importance to those detected by LIGO and/or VIRGO. In particular, there are several sources of cosmological origin of GWs that emit in the frequency range from few GHz to $10^{18}$ Hz that include primordial black holes, cosmic strings etc., see Refs. \cite{Dolgov:2011cq} and \cite{Ejlli:2019bqj} for more details.

The primary issue with GWs in the high frequency range is that at the moment there is not any direct way such that used by LIGO and VIRGO that could possibly detect them. The reason of such difficulty comes from the fact that at these high frequencies one must have extremely high sensitive ordinary GW detectors to achieve the predicted strain sensitivity. One way to pursue detection of GWs at these high frequencies is to use an indirect way such as the transformation of GWs into electromagnetic radiation in the presence of an external electromagnetic field. This interesting effect comes as a result of interaction of GWs with electromagnetic waves that produces electromagnetic waves out of GWs, see Ref. \cite{Ejlli:2018hke} for details. 

The transformation of GWs into electromagnetic waves, in the particular case, in the presence of an external magnetic field, has been studied in the literature by many authors and it has been applied in many contexts, see Refs. \cite{Ejlli:2018hke}-\cite{Dolgov:2012be}. However, one important aspect of all these studies is that the mixing and transformation of GWs into electromagnetic radiation has never been solved exactly in a constant external magnetic field. In Refs. \cite{Boccaletti70}- \cite{Zeldovich73} the graviton-photon mixing has been solved for those cases where back reaction effects and medium effects on electromagnetic radiation have not been considered. Furthermore, in those cases when medium and back reaction effects were included, only approximate solutions of the equations of motion were found. In these studies also some peculiar features of the propagating GW were found due to the not use of gauge invariant methods since these studies did not make use of the TT-gauge to remove unphysical degrees of freedom in the specific case of graviton-photon mixing. The latter case was included in the study of Ref. \cite{DeLogi77} where quantum field theory approach was used but no medium effects on the electromagnetic radiation were included. On the other hand, back reaction effects, correct gauge transformation and medium effects on the electromagnetic radiation were included in Ref. \cite{Raffelt88} but the equations of motion describing the mixing of GWs with electromagnetic radiation were solved only approximately by a WBK-like method of linearization of the equation of motion. Also a different field theory approach for the inverse process, namely of that of photon-graviton mixing has been studied in Ref. \cite{Bastianelli05}.

With the problems related to the theory of graviton-photon mixing mentioned above and with the fact that many sources of GWs emit in the very high frequency band of the spectrum and the only known way to detect them at present is by using the transformation of GWs into electromagnetic radiation, it is rather important to have exact solutions of the equations of motion and find exact expression for the intensity of produced electromagnetic radiation in a laboratory magnetic field. In this work, I address all the problems mentioned above and find for the first time in the literature exact solution of the equations of motion of graviton-photon mixing in a constant magnetic field and derive user friendly relations in the case when GWs are stochastic in nature. This paper is organized as follows: In Sec. \ref{sec:2}, I formulate the problem of graviton-photon mixing. In Sec. \ref{sec:3}, I find exact analytical expressions for the amplitudes of GWs and electromagnetic waves. In Sec. \ref{sec:4}, I derive important quantities of the produced electromagnetic radiation that can be compared with standard quantities of GW spectrum. In Sec. \ref{sec:5}, I conclude. In this work I use the metric with signature $\eta_{\mu\nu}=\text{diag}[1, -1, -1, -1]$ and work with the rationalized Lorentz-Heaviside natural units ($k_B=\hbar=c=\varepsilon_0=\mu_0=1$) with $e^2=4\pi \alpha$.

\section{Formulation of the problem}
\label{sec:2}

In order to formulate the problem that we are going to treat, let us start by expanding the metric tensor around the flat Minkowski spacetime as $g_{\mu\nu}=\eta_{\mu\nu}+\kappa h_{\mu\nu}+...$, and after  we get the following expression for the total effective action, see Ref. \cite{Ejlli:2018hke}: 
\begin{align}\label{tot-lang}
\mathcal S_\text{eff} & = \frac{1}{4}\int d^4 x \left[2\partial_\mu h^{\mu\nu}\partial_\rho h_\nu^\rho+\partial_\mu h\partial^\mu h-\partial_\mu h_{\alpha\beta}\partial^\mu h^{\alpha\beta}-2\partial_\mu h^{\mu\nu}\partial_\nu h\right]-\frac{1}{4}\int d^4 x F_{\mu\nu}F^{\mu\nu}+ \frac{\kappa}{2}\int d^4x h_{\mu\nu} T_\text{em}^{\mu\nu}\nonumber\\ & -\frac{1}{2}\int d^4 x\int d^4x^\prime A_\mu(x) \Pi^{\mu\nu}(x, x^\prime)A_\nu(x^\prime) + O(\kappa (\partial h)^3)+O(\kappa h_{\mu\nu} \Pi^{\mu\nu}),
\end{align}
where $h_{\mu\nu}$ is the gravitational wave tensor with $h=\eta_{\mu\nu}h^{\mu\nu}$, $\kappa^2\equiv 16\pi G_N$ where $G_N$ is the Newton constant and $T_\text{em}^{\mu\nu}$ is the electromagnetic field energy momentum tensor\footnote{With the metric with signature $\eta_{\mu\nu}=\text{diag}[1, -1, -1, -1]$, the expressions for the spatial components of the electromagnetic stress-energy tensor are $T_{ij}=\mathcal E_i \mathcal E_j + \mathcal B_i \mathcal B_j - (1/2)\delta_{ij}(\mathcal E^2+\mathcal B^2)$ where $\mathcal E_i=E_i+\bar E_i, \mathcal B_i=B_i+\bar B_i$ are respectively the components of the total electric and magnetic fields. In addition, a constant magnetic field itself is not a source of GWs since in order to generate GWs we must have spacetime varying fields.}. Here we are assuming that the medium where GWs and electromagnetic waves propagate is homogeneous where $\Pi^{\mu\nu}(x, x^\prime)= \Pi^{\mu\nu}(x-x^\prime)$ is the photon polarization tensor in a medium. Here we do not consider the effect that the medium has on the GWs due to their weak interaction with respect to the electromagnetic interaction.
Suppose that we have GWs propagating in vacuum and after they enter a region where only an external magnetic field is present. Since we consider the case when we are far away from the GW source(s), we can put GWs in the TT gauge before\footnote{Here we are imposing the gauge conditions for GWs already in the action and not after we find the equations of motion. Whether these two approaches give the same dynamics for the graviton-photon mixing is not clearly known and this issue is an open one and beyond the scope of this work. We expect that if other physical GW states are excited in addition to the usual GW states in vacuum, their impact on the whole dynamics is supposed to be very small due to the small amplitude of the original incident GW. } entering the magnetic field region, namely $h_{0\mu}=0, \partial^j h_{ij}=0,h_i^i=0$. 
In this case the effective action becomes 
\begin{equation}\label{tot-lang-1}
\mathcal S_\text{eff}  = -\frac{1}{4}\int d^4 x \partial_\mu h_{ij}\partial^\mu h^{ij}-\frac{1}{4}\int d^4 x F_{\mu\nu}F^{\mu\nu}+ \frac{\kappa}{2}\int d^4x h_{ij} (\mathcal E^i\mathcal E^j + \mathcal B^i\mathcal B^j) -\frac{1}{2}\int d^4 x\int d^4x^\prime A_\mu(x) \Pi^{\mu\nu}(x- x^\prime)A_\nu(x^\prime).
\end{equation}

The Euler-Lagrange equations of motion from the action in \eqref{tot-lang-1} for the propagating electromagnetic and gravitational fields components, $A^\mu$ and $h_{ij}$, propagating in the external constant and uniform magnetic field, are given by 
\begin{eqnarray}
\nabla^2 A^0 &=& 0 ,\nonumber\\
 \Box \bs A^i +\left(\int d^4x^\prime \Pi^{i j}(x-x^\prime)\bs A_j(x^\prime)\right)+ \partial^i\partial_\mu A^\mu &=&  - \kappa\,(\partial_j h^{ik})\bar F_{k}^j,\nonumber\\
\Box h_{ij} &=&-\kappa\left(B_i\bar B_j+\bar B_i B_j \right), 
  \label{system2}
\end{eqnarray}
where we used the fact that $F_{\mu\nu}\tilde F^{\mu\nu}=-4 \bs E\cdot \bs B$, $\tilde F^{0i}=-\bs B^i$, the TT-gauge conditions and expanded $F_{\mu\nu}=f_{\mu\nu} + \bar F_{\mu\nu}$ where $f_{\mu\nu}$ is the propagating electromagnetic tensor (which is usually a small perturbation over the external electromagnetic field tensor) and $\bar F_{\mu\nu}$ is the external electromagnetic field tensor. 
We work in the Coulomb gauge where $\partial_i\bs A^i=0$ and choose from the first equation in system \eqref{system2} $A^0=0$. The latter condition makes possible that the Lorentz gauge reduces to the Coulomb gauge $\partial_\mu A^\mu=\partial_i\bs A^i=0$. Next we look for solutions of the equations of motion \eqref{system2} by expanding the fields $A_i(\bs x, t)$ and $h_{ij}(\bs x, t)$ as Fourier integrals
\begin{align}\label{field-expansion}
\bs A^i({\bs x}, t) &=\sum_{\lambda=x, y, z} \int_{-\infty}^{+\infty}\frac{d\omega}{2\pi} e_\lambda^{i}(\hat{\bs n}) A_{\lambda}({\bs x, \omega})e^{-i \omega t},\quad  h_{ij}(\bs x, t)=\sum_{\lambda^\prime=\times, +} \int_{-\infty}^{+\infty}\frac{d\omega}{2\pi}   h_{\lambda^\prime}(\bs x, \omega) \textrm{e}_{ij}^{\lambda^\prime}(\hat{\bs n}) e^{-i \omega t},
\end{align}
where $e_\lambda^i$ is the photon polarization vector, $e_{ij}^{\lambda^\prime}$ is the GW polarization tensor with $\lambda^\prime$ indicating the polarization index or helicity state, and $\hat{\bs n}$ is the direction of the propagation of the GW. Without any loss of generality, consider the case when GW propagates in a given coordinate system along the $z$ axis, namely $\hat{\bs n}=\hat{\bs z}$. In addition, we have that $A_{\lambda}({\bs x, \omega})= A_{\lambda}^*({\bs x, -\omega})$ and $h_{\lambda^\prime}(\bs x, \omega)= h_{\lambda^\prime}^*(\bs x, -\omega)$ in order to ensure that $\bs A^i({\bs x}, t) $ and $h_{ij}(\bs x, t)$ are real functions.

Consider now a constant and uniform external magnetic field with components $\bar{\bs B}(\bs x, t)=[\bar B_x, \bar B_y, \bar B_z]$ and the vector potential with components $\bs A(\bs x, t)=[A_x(\bs x, t), A_y(\bs x, t), A_z(\bs x, t)]$. With the GW and electromagnetic wave propagating along the $\hat{\bs z}$ axis, $h_{ij}=h_{ij}(z, t), \bs A_i=\bs A_i(z, t)$ and with the field expansion \eqref{field-expansion}, the equations of motion \eqref{system2} for the GW tensor $h_{ij}$ in terms of the GW polarization states $h_+$ and $h_\times$ are given by
\begin{eqnarray}\label{GW-eq}
\left[\omega^2+\partial_z^2\right] h_+(z, \omega) &=& -\kappa \left[\partial_z A_x (z, \omega) \bar B_y+\partial_z A_y(z, \omega) \bar B_x\right],\nonumber\\
\left[\omega^2+\partial_z^2\right] h_\times (z, \omega) &=& \kappa \left[\partial_z A_x(z, \omega) \bar B_x-\partial_z A_y(z, \omega) \bar B_y\right],
\end{eqnarray}
and the equations of motion for the photon field $\bs A$ components in \eqref{system2}, are given by
\begin{eqnarray}\label{PH-eq}
\left[ \omega^2 + \partial_z^2 - \Pi_{xx}(\omega) \right]A_x(z, \omega) - \Pi_{xy}(\omega)A_y(z, \omega) - \Pi_{xz}(\omega) A_z(z, \omega) &=&  \kappa \left[\partial_z h_+(z, \omega) \bar B_y-\partial_z h_\times (z, \omega) \bar B_x\right],\nonumber\\
\left[ \omega^2 + \partial_z^2 - \Pi_{yy}(\omega) \right]A_y (z, \omega) - \Pi_{yx}(\omega)A_x(z, \omega) - \Pi_{yz}(\omega) A_z(z, \omega) &=&  \kappa \left[\partial_z h_\times (z, \omega) \bar B_y + \partial_z h_+ (z, \omega) \bar B_x\right],\nonumber\\
\left[ \omega^2\delta_{zj} - \Pi_{z j}(\omega)\right] A_j(z, \omega) &=& 0,
\end{eqnarray}
where in the Coulomb gauge there is no propagating longitudinal electromagnetic wave $\partial_z A_z(z, t)=0$ and $\Pi^{ij}=\Pi_{ij}=\Pi_{ij}(\omega)$ are the elements of the photon polarization tensor of a homogeneous and uniform medium calculated in the local limit $\bs x^\prime\rightarrow \bs x$. We may note that the third equation in the system \eqref{PH-eq} is actually a constraint on $A_z$ \cite{Ejlli:2018hke}. We stress that in deriving the equations of motions \eqref{GW-eq} and \eqref{PH-eq} we have assumed that both electromagnetic and GW fields depend only on $z$ and not $x, y$. We will discuss later about this assumption. We may also note from \eqref{GW-eq} that the polarization states of GWs $h_{\times, +}$ do not mix with $\bar B_z$.

In order to solve the system of equations \eqref{GW-eq} and \eqref{PH-eq}, we can reduce these systems to first order systems by defining $h_+(z, \omega) \equiv x_1(z, \omega), \partial_z h_+(z, \omega) \equiv x_2(z, \omega), h_\times(z, \omega) \equiv x_3(z, \omega), \partial_z h_\times(z, \omega) \equiv x_4(z, \omega),   A_x(z, \omega) \equiv x_5(z, \omega),  \partial_z A_x(z, \omega) \equiv x_6(z, \omega), A_y(z, \omega) \equiv x_7(z, \omega),  \partial_z A_y(z, \omega) \equiv x_8(z, \omega)$. With these substitutions we get the following system of first order partial differential equations:
\begin{eqnarray}\label{sys-sem}
\partial_z x_1 (z, \omega)&=& x_2(z, \omega), \quad \partial_z x_2 (z, \omega) = -\omega^2 x_1(z, \omega) -\kappa\bar B_y x_6(z, \omega) - \kappa\bar B_x x_8(z, \omega), \quad \partial_z x_3 (z, \omega) = x_4(z, \omega), \nonumber\\
\partial_z x_4 (z, \omega)&=& -\omega^2 x_3(z, \omega) + \kappa\bar B_x x_6(z, \omega) - \kappa\bar B_y x_8(z, \omega), \quad \partial_z x_5 (z, \omega) = x_6(z, \omega), \nonumber\\
\partial_z x_6 (z, \omega)&=& \left[ \Pi_{xx}(\omega) -\omega^2\right] x_5(z, \omega) + \Pi_{xy}(\omega) x_7(z, \omega) + \kappa\bar B_y x_2(z, \omega) -\kappa\bar B_x x_4(z, \omega), \quad \partial_z x_7 (z, \omega) =  x_8(z, \omega), \nonumber\\
\partial_z x_8 (z, \omega)&=& \left[ \Pi_{yy}(\omega) -\omega^2\right] x_7(z, \omega) + \Pi_{yx}(\omega) x_5(z, \omega) + \kappa\bar B_y x_4(z, \omega) + \kappa\bar B_x x_2(z, \omega).
\end{eqnarray}
Now we define the column field  $X(z, \omega) \equiv \left[ x_1(z, \omega) , x_2(z, \omega) , x_3(z, \omega) , x_4(z, \omega) , x_5(z, \omega) , x_6(z, \omega) , x_7(z, \omega) ,\\ x_8(z, \omega)  \right]^{\text{T}}$ and the system of differential equations \eqref{sys-sem} can be written as
\begin{equation}\label{system-1}
\partial_z X(z, \omega) = A (\omega, \bar B_x, \bar B_y)X(z, \omega),
\end{equation}
where the matrix $A$ is given by
\begin{equation}\label{matrix-A}
A(\omega,  \bar B_x, \bar B_y)=\begin{pmatrix}
0 & 1 & 0 & 0 & 0 & 0 & 0 & 0\\
-\omega^2 & 0 & 0 & 0 & 0 & -\kappa\bar B_y & 0 & -\kappa\bar B_x\\
0 & 0 & 0 & 1 & 0 & 0 & 0 & 0\\
0 & 0 & -\omega^2 & 0 & 0 & \kappa\bar B_x & 0 & -\kappa\bar B_y\\
0 & 0 & 0 & 0 & 0 & 1 & 0 & 0 \\
0 & \kappa\bar B_y & 0 & -\kappa\bar B_x  & \Pi_{xx}(\omega) -\omega^2 & 0 &  \Pi_{xy}(\omega) & 0 \\
0 & 0 & 0 & 0 & 0 & 0 & 0 & 1 \\
0 &  \kappa\bar B_x & 0 & \kappa\bar B_y &\Pi_{yx}(\omega) & 0 & \Pi_{yy}(\omega) -\omega^2 & 0 \\
   \end{pmatrix}.      
\end{equation}
The system \eqref{system-1} in order to be solved has to be supplied with appropriate initial conditions and/or boundary conditions if needed.

\section{Exact analytical solution of equations of motion}
\label{sec:3}

Equations \eqref{system-1} are a system of linear differential equations with constant coefficients and in principle can be solved exactly.  This is due to the fact that the matrix $A$ is constant since it does not depend on the position $z$ and it commutes with itself for $z^\prime\neq z$, $[A, A]=0$. Consequently the general solution of \eqref{system-1} is given by
\begin{equation}
X(z, \omega) = \exp{\left[\int_{z_i}^z dz^\prime A(\omega,  \bar B_x, \bar B_y)\right]} X(z_i, \omega),
\end{equation}
where $ X(z_i, \omega)$ is the field defined above at the initial position $z_i$ where the interaction with the magnetic field starts. To find the exponential of the matrix $A$ is quite difficult since it involves very tedious and long calculations\footnote{Here I assume that the reader knows how to calculate the exponential of a given matrix. In the specific case of the matrix $A$ in \eqref{matrix-A} it is important to note that this matrix has 8 different eigenvalues and 8 linearly independent eigenvectors and consequently it is diagonalizable. Based on this fact, the exponential of $A$ can be calculated by $\exp[A]=P \exp[D] P^{-1}$ where $P$ is an invertible matrix formed with the eigenvectors of $A$ and $D$ is the diagonal form of $A$ formed with the eigenvalues of $A$. The matrix $A$ is invertible if det$[A]\neq 0$ which in the case considered in this work is valid only for $\omega^2>\Pi_{xx, yy}$ and $\omega>0$. }. Due to this fact, we can simplify our calculations by assuming that the magnetic field is completely transversal with respect to direction of propagation of GWs. This particular type of configuration can be easily achieved in a laboratory when one is free to choose at convenience the direction of the external magnetic field. So, we choose $\bar{\bs B}(\bs x, t)=[\bar B_x, 0, 0]$ which also implies that in the case when there is a medium in the presence of the magnetic field, only the diagonal elements of $\Pi_{ij}$ are different from zero since the Faraday effect is absent for traverse magnetic field, see Ref. \cite{Ejlli:2018hke} for details. In this case we have that $\Pi_{xy}=0=\Pi_{yx}$ and $\kappa\bar B_y=0$ in the matrix $A$ in \eqref{matrix-A}. Let us define for future convenience
\begin{equation}
\mathcal C_{\pm}^{x, y}\equiv \sqrt{-2\omega^2 - \kappa^2\bar B_x^2 + \Pi_{xx, yy} \pm \mathcal B^{x, y}}, \quad \mathcal B^{x, y} \equiv \sqrt{\kappa^4\bar B_x^4 + 2 \kappa^2\bar B_x^2 (2\omega^2-\Pi_{xx, yy}) + \Pi_{xx, yy}^2},
\end{equation}
where the labels $x$ and $y$ in $\mathcal C_{\pm}^{x, y}, \mathcal B^{x, y}$ are respectively related to the components of $\Pi_{xx}$ and $\Pi_{yy}$ of the photon polarization tensor.
After by taking the exponential of $A$ in the case of transverse magnetic field along $\hat{\bs x}$ direction, we get the following exact solution of the components of the field $X(z, \omega)$:
\begin{align}
x_1(z, \omega) &= \left\{ \frac{1}{2}  \left[ \cosh{\left(\frac{z\, \mathcal C_{-}^y }{\sqrt{2}} \right)} + \cosh{\left(\frac{z\, \mathcal C_{+}^y }{\sqrt{2}} \right)} \right]- \left(  \frac{\kappa^2\bar B_x^2 - \Pi_{yy}}{2 \mathcal B^y} \right) \left[  \cosh{\left( \frac{z\, \mathcal C_{-}^y }{\sqrt{2}}\right)} - \cosh{\left( \frac{z\, \mathcal C_{+}^y }{\sqrt{2}}\right)}\right]\right\} x_1(0, \omega)\nonumber \\
& - \frac{1}{2\sqrt{2} \mathcal B^y} \left\{ \frac{e^{-z \mathcal C_-^y/\sqrt{2}}\left( \kappa^2\bar B_x^2 + \Pi_{yy} + \mathcal B^y\right)}{\mathcal C_-^y}  - \frac{e^{-z \mathcal C_+^y/\sqrt{2}}\left( \kappa^2\bar B_x^2 + \Pi_{yy} - \mathcal B^y\right)}{\mathcal C_+^y}  +   \frac{e^{z \mathcal C_+^y/\sqrt{2}}\left( \kappa^2\bar B_x^2 + \Pi_{yy} - \mathcal B^y\right)}{\mathcal C_+^y}  \right. \nonumber\\ & \left. -    \frac{e^{z \mathcal C_-^y/\sqrt{2}}\left( \kappa^2\bar B_x^2 + \Pi_{yy} + \mathcal B^y\right)}{\mathcal C_-^y}  \right\} x_2(0, \omega)    +    \frac{\sqrt{2} \kappa\bar B_x (\omega^2 -\Pi_{yy})}{\mathcal B^y \mathcal C_+^y \mathcal C_-^y} \left\{  \sinh{\left( \frac{z\, \mathcal C_+^y}{\sqrt{2}}\right)}\, \mathcal C_-^y - \sinh{\left( \frac{z\, \mathcal C_-^y}{\sqrt{2}}\right)}\, \mathcal C_+^y \right\} x_7(0, \omega) \nonumber\\
& + \frac{\kappa\bar B_x}{\mathcal B^y}  \left\{ \cosh{\left( \frac{z\, \mathcal C_-^y}{\sqrt{2}}\right)} - \cosh{\left( \frac{z\, \mathcal C_+^y}{\sqrt{2}}\right)} \right\} x_8(0, \omega), \nonumber\\
x_2(z, \omega) & = \frac{\omega^2}{2\sqrt{2} \mathcal B^y} \left\{ \frac{e^{z \mathcal C_+^y/\sqrt{2}}\left( \kappa^2\bar B_x^2 + \Pi_{yy} - \mathcal B^y\right)}{\mathcal C_+^y}  - \frac{e^{-z \mathcal C_+^y/\sqrt{2}}\left( \kappa^2\bar B_x^2 + \Pi_{yy} - \mathcal B^y\right)}{\mathcal C_+^y}  +   \frac{e^{- z \mathcal C_-^y/\sqrt{2}}\left( \kappa^2\bar B_x^2 + \Pi_{yy} + \mathcal B^y\right)}{\mathcal C_-^y}  \right. \nonumber\\ & \left. -    \frac{e^{z \mathcal C_-^y/\sqrt{2}}\left( \kappa^2\bar B_x^2 + \Pi_{yy} + \mathcal B^y\right)}{\mathcal C_-^y}  \right\} x_1(0, \omega)  +  \left\{\frac{1}{2} \left[ \cosh{\left( \frac{z\, \mathcal C_-^y}{\sqrt{2}}\right)} + \cosh{\left( \frac{z\, \mathcal C_+^y}{\sqrt{2}}\right)} \right] +\right. \nonumber\\ 
&\left.  \left(  \frac{\kappa^2\bar B_x^2 + \Pi_{yy}}{2 \mathcal B^y} \right) \left[  \cosh{\left( \frac{z\, \mathcal C_-^y}{\sqrt{2}}\right)}  - \cosh{\left( \frac{z\, \mathcal C_+^y}{\sqrt{2}}\right)} \right]\right\} x_2(0, \omega) - \frac{\kappa\bar B_x (\omega^2 -\Pi_{yy})}{\mathcal B^y} \left\{ \cosh{\left( \frac{z\, \mathcal C_-^y}{\sqrt{2}}\right)} - \cosh{\left( \frac{z\, \mathcal C_+^y}{\sqrt{2}}\right)} \right\} x_7(0, \omega)\nonumber\\
& +  \frac{ \kappa\bar B_x }{\sqrt{2} \mathcal B^y } \left\{  \sinh{\left( \frac{z\, \mathcal C_-^y}{\sqrt{2}}\right)}\, \mathcal C_-^y  - \sinh{\left( \frac{z\, \mathcal C_+^y}{\sqrt{2}}\right)}\, \mathcal C_+^y \right\} x_8(0, \omega), \nonumber
\end{align}
\begin{align}
x_3(z, \omega) & = \left\{ \frac{1}{2} \left[ \cosh{\left(\frac{z\, \mathcal C_{-}^x }{\sqrt{2}} \right)} + \cosh{\left(\frac{z\, \mathcal C_{+}^x }{\sqrt{2}} \right)} \right]- \left(  \frac{\kappa^2\bar B_x^2 - \Pi_{xx}}{ 2\mathcal B^x} \right) \left[  \cosh{\left( \frac{z\, \mathcal C_{-}^x }{\sqrt{2}}\right)} - \cosh{\left( \frac{z\, \mathcal C_{+}^x }{\sqrt{2}}\right)}\right]\right\} x_3(0, \omega)\, + \nonumber\\
& \frac{1}{2\sqrt{2} \mathcal B^x} \left\{ \frac{e^{- z \mathcal C_+^x/\sqrt{2}}\left( \kappa^2\bar B_x^2 + \Pi_{xx} - \mathcal B^x\right)}{\mathcal C_+^x}  - \frac{e^{z \mathcal C_+^x/\sqrt{2}}\left( \kappa^2\bar B_x^2 + \Pi_{xx} - \mathcal B^x\right)}{\mathcal C_+^x}  +   \frac{e^{ z \mathcal C_-^x/\sqrt{2}}\left( \kappa^2\bar B_x^2 + \Pi_{xx} + \mathcal B^x\right)}{\mathcal C_-^x}  \right. \nonumber\\ & \left. -    \frac{e^{- z \mathcal C_-^x/\sqrt{2}}\left( \kappa^2\bar B_x^2 + \Pi_{xx} + \mathcal B^x\right)}{\mathcal C_-^x}  \right\} x_4(0, \omega) - \frac{\sqrt{2} \kappa\bar B_x (\omega^2 -\Pi_{xx})}{\mathcal B^x \mathcal C_+^x \mathcal C_-^x } \left\{  \sinh{\left( \frac{z\, \mathcal C_+^x}{\sqrt{2}}\right)}\, \mathcal C_-^x - \sinh{\left( \frac{z\, \mathcal C_-^x}{\sqrt{2}}\right)}\, \mathcal C_+^x \right\} x_5(0, \omega) \nonumber\\
 &+ \frac{\kappa\bar B_x}{\mathcal B^x}  \left\{ \cosh{\left( \frac{z\, \mathcal C_+^x}{\sqrt{2}}\right)} - \cosh{\left( \frac{z\, \mathcal C_-^x}{\sqrt{2}}\right)} \right\} x_6(0, \omega),\nonumber\\
 x_4(z, \omega) &=  \frac{\omega^2}{2\sqrt{2} \mathcal B^x} \left\{ \frac{e^{ z \mathcal C_+^x/\sqrt{2}}\left( \kappa^2\bar B_x^2 + \Pi_{xx} - \mathcal B^x\right)}{\mathcal C_+^x}  - \frac{e^{- z \mathcal C_+^x/\sqrt{2}}\left( \kappa^2\bar B_x^2 + \Pi_{xx} - \mathcal B^x\right)}{\mathcal C_+^x}  +   \frac{e^{ - z \mathcal C_-^x/\sqrt{2}}\left( \kappa^2\bar B_x^2 + \Pi_{xx} + \mathcal B^x\right)}{\mathcal C_-^x}  \right. \nonumber\\ & \left. -    \frac{e^{ z \mathcal C_-^x/\sqrt{2}}\left( \kappa^2\bar B_x^2 + \Pi_{xx} + \mathcal B^x\right)}{\mathcal C_-^x}  \right\} x_3(0, \omega) \, + \nonumber
 \end{align}
 \vspace{-0.5cm}
\begin{align}
& \left\{  \frac{1}{2} \left[ \cosh{\left(\frac{z\, \mathcal C_{-}^x }{\sqrt{2}} \right)} + \cosh{\left(\frac{z\, \mathcal C_{+}^x }{\sqrt{2}} \right)} \right]+ \left(  \frac{\kappa^2\bar B_x^2 + \Pi_{xx}}{ 2\mathcal B^x} \right) \left[  \cosh{\left( \frac{z\, \mathcal C_{-}^x }{\sqrt{2}}\right)} - \cosh{\left( \frac{z\, \mathcal C_{+}^x }{\sqrt{2}}\right)}\right]\right\} x_4(0, \omega) + \nonumber\\
& \left[  \frac{\kappa\bar B_x(\omega^2- \Pi_{xx})}{ \mathcal B^x} \right] \left[  \cosh{\left( \frac{z\, \mathcal C_{-}^x }{\sqrt{2}}\right)} - \cosh{\left( \frac{z\, \mathcal C_{+}^x }{\sqrt{2}}\right)}\right] x_5(z, \omega) + \frac{\kappa\bar B_x}{\mathcal B^x } \left\{  \sinh{\left( \frac{z\, \mathcal C_+^x}{\sqrt{2}}\right)}\, \mathcal C_+^x - \sinh{\left( \frac{z\, \mathcal C_-^x}{\sqrt{2}}\right)}\, \mathcal C_-^x \right\} x_6(0, \omega), \nonumber\\
x_5(z, \omega) & = \frac{\sqrt{2}\omega^2\kappa\bar B_x}{\mathcal B^x } \left\{  \sinh{\left( \frac{z\, \mathcal C_+^x}{\sqrt{2}}\right)}/ \mathcal C_+^x - \sinh{\left( \frac{z\, \mathcal C_-^x}{\sqrt{2}}\right)}/ \mathcal C_-^x \right\} x_3(0, \omega) + \left[  \frac{\kappa\bar B_x}{ \mathcal B^x} \right] \left[  \cosh{\left( \frac{z\, \mathcal C_{-}^x }{\sqrt{2}}\right)} - \cosh{\left( \frac{z\, \mathcal C_{+}^x }{\sqrt{2}}\right)}\right] x_4(z, \omega) \nonumber\\
& + \left\{ \frac{1}{2} \left[ \cosh{\left(\frac{z\, \mathcal C_{-}^x }{\sqrt{2}} \right)} + \cosh{\left(\frac{z\, \mathcal C_{+}^x }{\sqrt{2}} \right)} \right]- \left(  \frac{\kappa^2\bar B_x^2 + \Pi_{xx}}{ 2\mathcal B^x} \right) \left[  \cosh{\left( \frac{z\, \mathcal C_{-}^x }{\sqrt{2}}\right)} - \cosh{\left( \frac{z\, \mathcal C_{+}^x }{\sqrt{2}}\right)}\right]\right\} x_5(0, \omega)\nonumber\\
&+ \frac{1}{2\sqrt{2} \mathcal B^x} \left\{ \frac{e^{ -z \mathcal C_-^x/\sqrt{2}}\left( - \kappa^2\bar B_x^2 + \Pi_{xx} - \mathcal B^x\right)}{\mathcal C_-^x}  + \frac{e^{ z \mathcal C_-^x/\sqrt{2}}\left( \kappa^2\bar B_x^2 - \Pi_{xx} + \mathcal B^x\right)}{\mathcal C_-^x}  +   \frac{e^{ - z \mathcal C_+^x/\sqrt{2}}\left( \kappa^2\bar B_x^2 - \Pi_{xx} - \mathcal B^x\right)}{\mathcal C_+^x}  \right. \nonumber\\ & \left. -    \frac{e^{ z \mathcal C_+^x/\sqrt{2}}\left( \kappa^2\bar B_x^2 - \Pi_{xx} - \mathcal B^x\right)}{\mathcal C_+^x}  \right\} x_6(0, \omega),\nonumber\\
x_6(z, \omega) & =  \left[  \frac{\omega^2\kappa\bar B_x}{ \mathcal B^x} \right] \left[  \cosh{\left( \frac{z\, \mathcal C_{+}^x }{\sqrt{2}}\right)} - \cosh{\left( \frac{z\, \mathcal C_{-}^x }{\sqrt{2}}\right)}\right] x_3(z, \omega) + \frac{\kappa\bar B_x}{\sqrt{2} \mathcal B^x } \left\{  \sinh{\left( \frac{z\, \mathcal C_-^x}{\sqrt{2}}\right)}\,\mathcal C_-^x - \sinh{\left( \frac{z\, \mathcal C_+^x}{\sqrt{2}}\right)}\,\mathcal C_+^x \right\} x_4(0, \omega)\nonumber\\
& + \frac{\omega^2-\Pi_{xx}}{ 2\sqrt{2}\mathcal B^x} \left\{ \frac{e^{ -z \mathcal C_-^x/\sqrt{2}}\left(  \kappa^2\bar B_x^2 - \Pi_{xx} + \mathcal B^x\right)}{\mathcal C_-^x}  - \frac{e^{ z \mathcal C_-^x/\sqrt{2}}\left( \kappa^2\bar B_x^2 - \Pi_{xx} + \mathcal B^x\right)}{\mathcal C_-^x}  +   \frac{e^{ - z \mathcal C_+^x/\sqrt{2}}\left( -\kappa^2\bar B_x^2 + \Pi_{xx} + \mathcal B^x\right)}{\mathcal C_+^x}  \right. \nonumber\\ & \left. -    \frac{e^{ z \mathcal C_+^x/\sqrt{2}}\left(- \kappa^2\bar B_x^2 + \Pi_{xx} + \mathcal B^x\right)}{\mathcal C_+^x}  \right\} x_5(0, \omega) +  \left\{ \frac{1}{2} \left[ \cosh{\left(\frac{z\, \mathcal C_{-}^x }{\sqrt{2}} \right)} + \cosh{\left(\frac{z\, \mathcal C_{+}^x }{\sqrt{2}} \right)} \right]+ \right.\nonumber\\ & \left. \left(  \frac{\kappa^2\bar B_x^2 - \Pi_{xx}}{ 2\mathcal B^x} \right) \left[  \cosh{\left( \frac{z\, \mathcal C_{-}^x }{\sqrt{2}}\right)} - \cosh{\left( \frac{z\, \mathcal C_{+}^x }{\sqrt{2}}\right)}\right]\right\} x_6(0, \omega),\nonumber
\end{align}
\begin{align}\label{solutions-1}
x_7(z, \omega) &= \frac{\sqrt{2}\omega^2\kappa\bar B_x}{\mathcal B^y } \left\{  \sinh{\left( \frac{z\, \mathcal C_-^y}{\sqrt{2}}\right)}/ \mathcal C_-^y - \sinh{\left( \frac{z\, \mathcal C_+^y}{\sqrt{2}}\right)}/ \mathcal C_+^y \right\} x_1(0, \omega) + \left[  \frac{\kappa\bar B_x}{ \mathcal B^y} \right] \left[  \cosh{\left( \frac{z\, \mathcal C_{+}^y }{\sqrt{2}}\right)} - \cosh{\left( \frac{z\, \mathcal C_{-}^y }{\sqrt{2}}\right)}\right] x_2(z, \omega) \nonumber\\
& +  \left\{ \frac{1}{2} \left[ \cosh{\left(\frac{z\, \mathcal C_{-}^y }{\sqrt{2}} \right)} + \cosh{\left(\frac{z\, \mathcal C_{+}^y }{\sqrt{2}} \right)} \right]- \left(  \frac{\kappa^2\bar B_x^2 + \Pi_{yy}}{ 2\mathcal B^y} \right) \left[  \cosh{\left( \frac{z\, \mathcal C_{-}^y }{\sqrt{2}}\right)} - \cosh{\left( \frac{z\, \mathcal C_{+}^y }{\sqrt{2}}\right)}\right]\right\} x_7(0, \omega)\, + \nonumber\\
& \frac{1}{2\sqrt{2} \mathcal B^y} \left\{ \frac{e^{ -z \mathcal C_-^y/\sqrt{2}}\left( - \kappa^2\bar B_x^2 + \Pi_{yy} - \mathcal B^y\right)}{\mathcal C_-^y}  + \frac{e^{ z \mathcal C_-^y/\sqrt{2}}\left( \kappa^2\bar B_x^2 - \Pi_{yy} + \mathcal B^y\right)}{\mathcal C_-^y}  +   \frac{e^{ - z \mathcal C_+^y/\sqrt{2}}\left( \kappa^2\bar B_x^2 - \Pi_{yy} - \mathcal B^y\right)}{\mathcal C_+^y}  \right. \nonumber\\ & \left. -    \frac{e^{ z \mathcal C_+^y/\sqrt{2}}\left( \kappa^2\bar B_x^2 - \Pi_{yy} - \mathcal B^y\right)}{\mathcal C_+^y}  \right\} x_8(0, \omega),\nonumber\\
x_8(z, \omega) &= \left[  \frac{\kappa\omega^2 \bar B_x}{ \mathcal B^y} \right] \left[  \cosh{\left( \frac{z\, \mathcal C_{-}^y }{\sqrt{2}}\right)} - \cosh{\left( \frac{z\, \mathcal C_{+}^y }{\sqrt{2}}\right)}\right] x_1(z, \omega) +  \frac{\kappa\bar B_x}{\sqrt{2}\mathcal B^y } \left\{  \sinh{\left( \frac{z\, \mathcal C_+^y}{\sqrt{2}}\right)}\, \mathcal C_+^y - \sinh{\left( \frac{z\, \mathcal C_-^y}{\sqrt{2}}\right)}\, \mathcal C_-^y \right\} x_2(0, \omega)\, + \nonumber\\
& \frac{\omega^2-\Pi_{yy}}{2\sqrt{2} \mathcal B^y} \left\{ \frac{e^{ -z \mathcal C_-^y/\sqrt{2}}\left(  \kappa^2\bar B_x^2 - \Pi_{yy} + \mathcal B^y\right)}{\mathcal C_-^y}  - \frac{e^{ z \mathcal C_-^y/\sqrt{2}}\left( \kappa^2\bar B_x^2 - \Pi_{yy} + \mathcal B^y\right)}{\mathcal C_-^y}  -   \frac{e^{ - z \mathcal C_+^y/\sqrt{2}}\left( \kappa^2\bar B_x^2 - \Pi_{yy} - \mathcal B^y\right)}{\mathcal C_+^y}  \right. \nonumber\\ & \left. -    \frac{e^{ z \mathcal C_+^y/\sqrt{2}}\left( \kappa^2\bar B_x^2 - \Pi_{yy} - \mathcal B^y\right)}{\mathcal C_+^y}  \right\} x_7(0, \omega)\nonumber\\
& +  \left\{ \frac{1}{2} \left[ \cosh{\left(\frac{z\, \mathcal C_{-}^y }{\sqrt{2}} \right)} + \cosh{\left(\frac{z\, \mathcal C_{+}^y }{\sqrt{2}} \right)} \right]+ \left(  \frac{\kappa^2\bar B_x^2 - \Pi_{yy}}{ 2\mathcal B^y} \right) \left[  \cosh{\left( \frac{z\, \mathcal C_{-}^y }{\sqrt{2}}\right)} - \cosh{\left( \frac{z\, \mathcal C_{+}^y }{\sqrt{2}}\right)}\right]\right\} x_8(0, \omega),
\end{align}
where for simplicity we have chosen at $z_i=0$ the origin of our coordinate system.

With the solutions in \eqref{solutions-1} we have all necessary quantities to find the exact solutions of the original graviton and photon states $h_{+, \times}$ and $A_{x, y}$. We may see from \eqref{solutions-1} that the solutions of the original fields $h_{+, \times}(z, \omega)= x_{1, 3}(z, \omega)$ are proportional to the field derivatives $\partial_z h_{+, \times}(0, \omega)$ and $\partial _zA_{x, y}(0, \omega)$ as one should expect from the initial second order system of differential equations \eqref{GW-eq} and \eqref{PH-eq}. Since we are interested in the generation of electromagnetic radiation in the laboratory, we have that $A_{x, y}(z=0, \omega)=0$ and $\partial_z A_{x, y}(z=0, \omega)=0$ which are our initial conditions. With these considerations and by replacing $\tilde h_{ij}(\bs x, t)=\kappa h_{ij}(\bs x, t)$ with $\tilde h_{ij}(\bs x, t)$ being the dimensionless GW components, we get from \eqref{solutions-1}
\begin{align}\label{solutions-2}
\tilde h_+(z, \omega) &= \left\{\frac{1}{2}  \left[ \cosh{\left(\frac{z\, \mathcal C_{-}^y }{\sqrt{2}} \right)} + \cosh{\left(\frac{z\, \mathcal C_{+}^y }{\sqrt{2}} \right)}\right] - \left(  \frac{\kappa^2\bar B_x^2 - \Pi_{yy}}{2 \mathcal B^y} \right) \left[  \cosh{\left( \frac{z\, \mathcal C_{-}^y }{\sqrt{2}}\right)} - \cosh{\left( \frac{z\, \mathcal C_{+}^y }{\sqrt{2}}\right)}\right]\right\} \tilde h_+(0, \omega)\nonumber \\
& - \frac{1}{2\sqrt{2} \mathcal B^y} \left\{ \frac{e^{-z \mathcal C_-^y/\sqrt{2}}\left( \kappa^2\bar B_x^2 + \Pi_{yy} + \mathcal B^y\right)}{\mathcal C_-^y}  - \frac{e^{-z \mathcal C_+^y/\sqrt{2}}\left( \kappa^2\bar B_x^2 + \Pi_{yy} - \mathcal B^y\right)}{\mathcal C_+^y}  +   \frac{e^{z \mathcal C_+^y/\sqrt{2}}\left( \kappa^2\bar B_x^2 + \Pi_{yy} - \mathcal B^y\right)}{\mathcal C_+^y}  \right. \nonumber\\ & \left. -    \frac{e^{z \mathcal C_-^y/\sqrt{2}}\left( \kappa^2\bar B_x^2 + \Pi_{yy} + \mathcal B^y\right)}{\mathcal C_-^y}  \right\} \partial_z \tilde h_+(0, \omega),\nonumber\\
\tilde h_\times(z, \omega) & = \left\{\frac{1}{2}  \left[ \cosh{\left(\frac{z\, \mathcal C_{-}^x }{\sqrt{2}} \right)} + \cosh{\left(\frac{z\, \mathcal C_{+}^x }{\sqrt{2}} \right)} \right]- \left(  \frac{\kappa^2\bar B_x^2 - \Pi_{xx}}{ 2\mathcal B^x} \right) \left[  \cosh{\left( \frac{z\, \mathcal C_{-}^x }{\sqrt{2}}\right)} - \cosh{\left( \frac{z\, \mathcal C_{+}^x }{\sqrt{2}}\right)}\right]\right\} \tilde h_\times(0, \omega)\, + \nonumber\\
& \frac{1}{2\sqrt{2} \mathcal B^x} \left\{ \frac{e^{- z \mathcal C_+^x/\sqrt{2}}\left( \kappa^2\bar B_x^2 + \Pi_{xx} - \mathcal B^x\right)}{\mathcal C_+^x}  - \frac{e^{z \mathcal C_+^x/\sqrt{2}}\left( \kappa^2\bar B_x^2 + \Pi_{xx} - \mathcal B^x\right)}{\mathcal C_+^x}  +   \frac{e^{ z \mathcal C_-^x/\sqrt{2}}\left( \kappa^2\bar B_x^2 + \Pi_{xx} + \mathcal B^x\right)}{\mathcal C_-^x}  \right. \nonumber\\ & \left. -    \frac{e^{- z \mathcal C_-^x/\sqrt{2}}\left( \kappa^2\bar B_x^2 + \Pi_{xx} + \mathcal B^x\right)}{\mathcal C_-^x}  \right\} \partial_z \tilde h_\times(0, \omega),\nonumber\\
A_x(z, \omega) & = \frac{\sqrt{2}\omega^2\bar B_x}{\mathcal B^x } \left\{  \sinh{\left( \frac{z\, \mathcal C_+^x}{\sqrt{2}}\right)}/ \mathcal C_+^x - \sinh{\left( \frac{z\, \mathcal C_-^x}{\sqrt{2}}\right)}/ \mathcal C_-^x \right\} \tilde h_\times(0, \omega) + \left[  \frac{\bar B_x}{ \mathcal B^x} \right] \left[  \cosh{\left( \frac{z\, \mathcal C_{-}^x }{\sqrt{2}}\right)} - \cosh{\left( \frac{z\, \mathcal C_{+}^x }{\sqrt{2}}\right)}\right] \partial_z \tilde h_\times(0, \omega),\nonumber\\
A_y(z, \omega) &= -\frac{\sqrt{2}\omega^2\bar B_x}{\mathcal B^y } \left\{  \sinh{\left( \frac{z\, \mathcal C_+^y}{\sqrt{2}}\right)}/ \mathcal C_+^y - \sinh{\left( \frac{z\, \mathcal C_-^y}{\sqrt{2}}\right)}/ \mathcal C_-^y \right\} \tilde h_+(0, \omega) - \left[  \frac{\bar B_x}{ \mathcal B^y} \right] \left[  \cosh{\left( \frac{z\, \mathcal C_{-}^y }{\sqrt{2}}\right)} - \cosh{\left( \frac{z\, \mathcal C_{+}^y }{\sqrt{2}}\right)}\right] \partial_z \tilde h_+(0, \omega).
\end{align}
One important thing to note about solutions \eqref{solutions-2} is that they are valid only in the case when exist a transverse magnetic field with non zero amplitude $\bar B_x\neq 0$ and for $\omega>0$. Another important fact to note about solutions \eqref{solutions-2} is that they depend on the derivative of the graviton fields at the initial position $z=0$, namely $\partial_z h_{+, \times}(z=0, \omega)$. A complete solutions is found by explicitly knowing the fields $h_{+, \times}(z, \omega)$ and their first order derivatives at $z=0$. Then with the expressions in \eqref{solutions-2} one can find the complete expressions for the fields in \eqref{field-expansion} by calculating the Fourier integrals. In case the fields are known to be monochromatic, then it is not necessary to calculate the Fourier integrals in \eqref{field-expansion}. However, even in the case when the fields are given as Fourier integrals, one can calculate them in some specific cases such as propagation in a vacuum. Even in the case when propagation is not in vacuum, for practical purposes we do not need to explicitly calculate the Fourier integrals because with will deal with quantities that are calculated per logarithmic frequency interval. We will see this in the next section.

\section{Energy flux and density parameter of generated electromagnetic radiation}
\label{sec:4}

While the solutions \eqref{solutions-2} are exact, it is very important as a matter of example, to consider some cases regarding the magnitude of some of the parameters that enter in \eqref{solutions-2}. First thing is to see if we can simplify and approximate the expressions for $\mathcal C_{\pm}^{x, y}$ and $\mathcal B^{x, y}$. By defining 
\begin{equation}\label{parameters}
\rho^2\equiv \frac{\kappa^2 \bar B_x^2}{2\omega^2}= 3.75 \times 10^{-30} \left( \frac{\text{Hz}}{f}\right)^2 \left( \frac{\bar B_x}{\text G}\right)^2, \quad \sigma_{xx, yy}^2 \equiv \frac{\Pi_{xx, yy}}{2\omega^2} = 2.92 \times 10^{28} \left( \frac{\text{Hz}}{f}\right)^2 \left( \frac{\Pi_{xx, yy}}{\text{eV}^2}\right),
\end{equation}
where $f=\omega/2\pi$ is the frequency of GWs and electromagnetic radiation, we can write $\mathcal C_{\pm}^{x, y}$ and $\mathcal B^{x, y}$ as
\begin{equation}\nonumber
\mathcal C_{\pm}^{x, y}= i \sqrt{2}\,\omega\, \left[ 1+ \rho^2 - \sigma_{xx, yy}^2 \mp \mathcal B^{x, y}/(2\omega^2) \right]^{1/2}, \quad \mathcal B^{x, y} /(2\omega^2)= \left[ \rho^4 + 2\rho^2 (1-\sigma_{xx, yy}^2) + \sigma_{xx, yy}^4\right]^{1/2}.
\end{equation}
The dimensionless parameter $\rho$ in \eqref{parameters} essentially gives the ratio of interaction energy of GWs with the magnetic field to the GW energy itself, while the dimensionless parameter $\sigma_{xx, yy}$ gives the ratio of photon effective mass in the medium to the photon energy. Values of $\rho\ll 1$ indicate that gravitons weakly interact with the medium and can be safely assumed relativistic. Indeed, as we can see from the definition of $\rho$ in \eqref{parameters}, for high frequency GWs and weak magnetic field, $\rho\ll 1$. On the other hand for very low GWs energies and strong magnetic fields, we can also have $\rho\gg 1$. Since in this work we are interested in very high frequency GWs and typical  laboratory magnetic fields, we have $\rho\ll 1$. The same arguments apply also to the parameter $\sigma_{xx, yy}$ which quantifies of how much relativistic are photons in the medium and more precisely it can be shown to be the deviation from unity of photon indexes of refraction in media. Since in most cases photons are relativistic in most laboratory media, we can take $\sigma_{xx, yy}\ll 1$. Based on these considerations we can write
\begin{equation}
\mathcal C_{\pm}^{x, y} \simeq i \sqrt{2}\,\omega\, \left[ 1+ \frac{1}{2}\left(\rho^2 - \sigma_{xx, yy}^2 \mp \mathcal B^{x, y}/(2\omega^2)\right) \right],
\end{equation}
and we get to first order the following expressions for the fields $A_{x, y}(z, \omega)$ in \eqref{solutions-2}
\begin{align}\label{solutions-3}
A_x(z, \omega) &=  \left\{\frac{\bar B_x}{2\omega} \sin\left[ \omega z \left( 1 + \frac{\rho^2 - \sigma_{xx}^2}{2}\right)\right] \cos\left[ \frac{ z \mathcal B^x}{4\omega}\right] - \frac{(2-\rho^2 + \sigma_{xx}^2)\omega\bar B_x}{\mathcal B^x} \cos\left[ \omega z \left( 1 + \frac{\rho^2 - \sigma_{xx}^2}{2}\right)\right] \sin\left[ \frac{z \mathcal B^x}{4\omega}\right] \right\} \tilde h_\times(0, \omega) \nonumber\\
& - \left[  \frac{2 \bar B_x}{ \mathcal B^x} \right] \sin\left[ \omega z \left( 1 + \frac{\rho^2 - \sigma_{xx}^2}{2}\right)\right] \sin\left[ \frac{ z \mathcal B^x}{4\omega}\right] \partial_z \tilde h_\times(0, \omega),\nonumber\\
A_y(z, \omega) &=  \left\{\frac{\bar B_x}{2\omega} \sin\left[ \omega z \left( 1 + \frac{\rho^2 - \sigma_{yy}^2}{2}\right)\right] \cos\left[ \frac{ z \mathcal B^y}{4\omega}\right] - \frac{(2-\rho^2 + \sigma_{yy}^2)\omega\bar B_x}{\mathcal B^y} \cos\left[ \omega z \left( 1 + \frac{\rho^2 - \sigma_{yy}^2}{2}\right)\right] \sin\left[ \frac{z \mathcal B^y}{4\omega}\right] \right\} \tilde h_+(0, \omega) \nonumber\\
& - \left[  \frac{2 \bar B_x}{ \mathcal B^y} \right] \sin\left[ \omega z \left( 1 + \frac{\rho^2 - \sigma_{yy}^2}{2}\right)\right] \sin\left[ \frac{ z \mathcal B^y}{4\omega}\right] \partial_z \tilde h_+(0, \omega).
\end{align}

We can use the solutions in \eqref{solutions-3} to discuss some particular cases that are of interest in many situations. The first thing is to see how the solutions \eqref{solutions-3} become in the WKB or Slowly Varying Envelope Approximation (SVEA) that has been widely used in the literature, see Refs. \cite{Ejlli:2019bqj} and \cite{Ejlli:2018hke} and other references therein. In the SVEA approximation one usually linearize the equation of motions \eqref{GW-eq} and \eqref{PH-eq} by two step related assumptions. One first writes the operator $(\omega^2 + \partial_z^2)(\cdot)=(\omega-i\partial_z)(\omega + i\partial_z)(\cdot)= (\omega +k)(\omega + i \partial_z)(\cdot)\simeq 2\omega (\omega + i \partial_z)(\cdot)$ by assuming that all mixing fields are relativistic with $k\simeq \omega$ or equivalently that the index of refraction of waves is very close to one. The symbol $(\cdot)$ stands for a general function of $z$ and $\omega$ such as the electromagnetic wave or GW amplitudes. The second assumption that is related to the fact that the fields are relativistic is by replacing spatial derivatives $\partial_z (\cdot)\simeq ik (\cdot)$, see Ref. \cite{Ejlli:2018hke} for more details. If we apply the SVEA approximations to the expressions \eqref{solutions-3}, we can neglect the terms $\rho^2$ and $\sigma_{xx, yy}^2$ with respect to unity. This is the relativistic approximation of the GW and electromagnetic field. The other approximation allows us to replace $\partial _z\tilde h_{\times, +}(z=0, \omega)\simeq ik \tilde h_{\times, +}(z=0, \omega)\simeq i\omega \tilde h_{\times, +}(z=0, \omega)$. If we make all these approximations, we find the same solutions as those found in Ref. \cite{Ejlli:2019bqj} (expressions B.8) for the fields $A_{x, y}(z, \omega)$.

A second thing to consider is to use solutions \eqref{solutions-3} to calculate for example the Stokes parameters and in particular the intensity (or energy flux) of the electromagnetic wave which is given by the Stokes parameter $I_\gamma$
\begin{equation}\label{Stokes}
I_\gamma(z, t)= \langle |E_x(z, t)|^2 \rangle +  \langle |E_y(z, t)|^2 \rangle,
\end{equation}
where $E_x$ and $E_y$ are the components of the electric field of the electromagnetic wave and the symbol $\langle (\cdot) \rangle$ denotes the temporal average of the electric field components over many oscillations periods. In Fourier space the time average is replaced by energy average over $\omega$. In addition, we have that $E_{x, y}(\bs x, t)= -\partial_t A_{x, y}(\bs x, t) =- i \int_{-\infty}^{+\infty} \frac{d\omega}{2\pi} \, \omega A_{x, y}(\bs x, \omega) e^{-i\omega t}$ for  $A^0(\bs x, t)=0$. In order to proceed further, we need to make some assumption about the nature of GWs that generate the electromagnetic radiation through the graviton-photon mixing. Usually, GWs that reach the Earth are from a single source such a binary system or a collection that are emitted by different uncorrelated sources such as a stochastic background. Let us consider for example the latter case in what follows. As shown in Ref.  \cite{Ejlli:2019bqj}, a stochastic background of GWs is assumed to be isotropic, stationary and unpolarized. Under such conditions, we have that the ensemble average of GW amplitudes satisfies
\begin{equation}\label{quant}
\langle \tilde h_\lambda(\hat{\bs n}, \omega) \tilde h_{\lambda^\prime}^*(\hat{\bs n}^\prime, \omega^\prime) \rangle = 2\pi \delta(\omega-\omega^\prime) \frac{\delta^2(\hat{\bs n}, \hat{\bs n}^\prime)}{4\pi}\,\delta_{\lambda\lambda^\prime} \frac{H(\omega)}{2},
\end{equation}
where $H(\omega)$ is the stochastic background spectral density that has the physical dimensions of Hz$^{-1}$ and $\delta^2(\hat{\bs n}, \hat{\bs n}^\prime)= \delta(\phi-\phi^\prime) \delta(\cos\theta-\cos\theta^\prime)$ is the covariant Dirac delta on the two sphere and $\phi, \theta$ are the usual spherical angular coordinates. Of course expression \eqref{quant} does not explicitly depend on the position $\bs x$.

Now it is more convenient to define the following functions
\begin{align}\nonumber
F_{x, y}(z, \omega, \bar B_x) & \equiv \frac{\bar B_x}{2\omega} \sin\left[ \omega z \left( 1 + \frac{\rho^2 - \sigma_{xx, yy}^2}{2}\right)\right] \cos\left[ \frac{ z \mathcal B^{x, y}}{4\omega}\right] - \frac{(2-\rho^2 + \sigma_{xx, yy}^2)\omega\bar B_x}{\mathcal B^{x, y}} \cos\left[ \omega z \left( 1 + \frac{\rho^2 - \sigma_{xx, yy}^2}{2}\right)\right] \sin\left[ \frac{z \mathcal B^{x, y}}{4\omega}\right],\nonumber\\
G_{x, y}(z, \omega, \bar B_x) & \equiv \left[  \frac{2 \bar B_x}{ \mathcal B^{x, y}} \right] \sin\left[ \omega z \left( 1 + \frac{\rho^2 - \sigma_{xx, yy}^2}{2}\right)\right] \sin\left[ \frac{ z \mathcal B^{x, y}}{4\omega}\right],\nonumber
\end{align}
where the subscripts $x, y$ in $F_{x, y}$ and $G_{x, y}$ indicate if do appear in them either $\sigma_{xx}$ and $\mathcal B^x$ or $\sigma_{yy}$ and $\mathcal B^y$. By using the definitions of  $F_{x, y}$ and $G_{x, y}$ above and making use of the ensemble average in \eqref{quant} we get the following expressions for $\langle |E_{x, y}(z, t)|^2 \rangle $:
\begin{equation}\label{electric-f}
\langle |E_{x, y}(z, t)|^2 \rangle = \int_{0}^{\infty} \frac{d\omega}{2\pi} \left[ | F_{x, y} |^2 + \omega^2 |G_{x, y}|^2 + 2\omega\, \text{Im} \{iF_{x, y} G_{x, y}^*\} \right] \omega^2 H(\omega),
\end{equation}
where we took $k=|\bs k| \simeq \pm \omega$ in deriving \eqref{electric-f} with $k$ being the module of the GW wave vector that lies along the $z$-axis in our case and used the fact that $\omega$ can have either sign depending on the integration interval. Relativistic gravitons have been assumed as discussed above and we used the fact that the integrand in \eqref{electric-f} is an even function in $\omega$. It is important to stress that the assumptions made for the stochastic background of GWs (unpolarized, isotropic and stationary) that start interacting with the magnetic field at $z=0$ are not directly assumptions on the electromagnetic field generated during the process of graviton-photon mixing. Indeed, the electromagnetic wave received at the detector is not isotropic but it comes from a specific direction, namely that along the $z$-axis.

For more practical uses we can use expressions \eqref{Stokes} and \eqref{electric-f} to calculate the energy density or intensity of the electromagnetic radiation per logarithmic energy interval. By using \eqref{electric-f}, we get the following expression
\begin{equation}\label{spectrum}
\frac{d I_\gamma(z, \omega)}{d(\log\omega)} =  \left[ | F_{x} |^2 +  | F_{y} |^2 + \omega^2\left(|G_{x}|^2 + |G_{y}|^2\right) +  2\omega\, \text{Im} \{i\left(F_{x} G_{x}^* + F_{y} G_{y}^*\right)\} \right] \frac{\omega^2 h_c^2(\omega)}{2},
\end{equation}
where $h_c^2(\omega)=\omega H(\omega)/\pi$ is defined as the characteristic amplitude of the stochastic background of GWs and it is a dimensionless quantity. Expression \eqref{spectrum} gives us the energy density of electromagnetic radiation generated by interaction of a stochastic background of GWs with a constant magnetic field. We can use expression \eqref{spectrum} to derive another important quantity which is the density parameter of electromagnetic waves generated through graviton-photon mixing in constant magnetic field. We recall that the density parameter is defined as $\Omega(z, \omega; t)\equiv (1/\rho_c)d\rho(z, \omega; t)/d(\log\omega)$ where $\rho$ is the energy density of a given field and $\rho_c= 6H_0^2/\kappa^2$ is the critical energy density of our Universe and $H_0= 100\,h_0$(km/s/Mpc) is the Hubble parameter with $h_0$ being a dimensionless parameter. At $z=0$ the total energy flux (or energy density) of GWs is given by
\begin{equation}\label{gw-intens}
I_\text{gw}(0, t)=\int_0^{+\infty} d(\log\omega) \frac{\omega^2 h_c^2(\omega)}{\kappa^2}.
\end{equation}
By using expressions \eqref{spectrum}-\eqref{gw-intens} and the definition of the density parameter $\Omega$, we get the following relation between the density parameters of electromagnetic waves at a distance $z$ from the origin with the density parameter of GWs at the origin $z=0$
 \begin{equation}\label{dens-par-gamma}
h_0^2\Omega_\gamma(z, \omega) = (\kappa^2/2) \left[ | F_{x} |^2 +  | F_{y} |^2 + \omega^2\left( |G_{x}|^2 + |G_{y}|^2\right) +  2\omega\, \text{Im} \{i\left(F_{x} G_{x}^* + F_{y} G_{y}^*\right)\} \right] h_0^2\Omega_\text{gw}(0, \omega).
\end{equation}

\section{Conclusions}
\label{sec:5}

In this work I studied and found exact solutions of the equations of motions of graviton-photon mixing in a constant and perpendicular magnetic field with respect to the direction of propagation of GWs. The results found in this work are new and improve previous results where approximate methods have been used to solve the equations of motion and possible medium effects on the electromagnetic waves were not included. The results found for the components of the electromagnetic field in \eqref{solutions-3} can be used to find characteristic quantities related to GWs such as the energy flux and the density parameter $h_0^2\Omega_\text{gw}$ and to make predictions that can be compared with experimental results as discussed in Ref. \cite{Ejlli:2019bqj}.

The solutions for the components of the electromagnetic field are different from those originally obtained by using approximative solutions that have been found in the SVEA or WKB approximation. As I have shown by exactly solving the equations of motion, there are extra terms when the full second order partial differential equations are solved with both initial conditions on the fields and their derivatives with respect to the position. In the case when the fields are relativistic and with slowly varying amplitudes within the SVEA approximation conditions, one recovers the approximate results previously found in the literature. With the full solution of the equations of motion, it is possible to calculate the expected density parameter of electromagnetic radiation generated in the graviton-photon mixing as obtained in expression 
\eqref{dens-par-gamma}. 

One important aspect that is worth to point out with the solutions \eqref{solutions-3} is that they have been obtained in the case when the propagation in the magnetic field does not happen within a confined and closed type detector such as a cavity. This fact is reflected on the assumptions that have been made on the graviton and photon fields that depend only on the coordinate $z$ along the direction of propagation. This assumption is correct when the detector is not closed and not confined in every direction such as a conducting cavity and the waves are expected to depend only on the propagation distance $z$ and do not depend on the $x, y$ coordinates. However, if the detector is confined in a fixed volume in every direction such as that of a conducting cavity, it is more appropriate to look for solutions of the electromagnetic field that depend on all coordinates $x, y, z$ instead of only $z$. This assumption is quite common when electromagnetic waves propagate in a cavity and are constantly reflected on its walls by forming within a short time standing waves. In such case one needs to solve the problem of graviton-photon mixing with the appropriate boundary conditions that reflect the form and shape of the cavity and the solutions found in \eqref{solutions-3} might not be appropriate for such type of detectors. Even though this is true, one might expect that even for a cavity type detector, solutions \eqref{solutions-3} might be sufficient to first order of approximation. This is analogous to the common electromagnetic cavities when typically a plane wave like solution of the electromagnetic field is used instead of a beam that depends on all coordinates $x, y, z$ and propagates in several directions within the cavity.


\begin{thebibliography}{99}



\bibitem{Dolgov:2011cq}
A.~D.~Dolgov and D.~Ejlli,
``Relic gravitational waves from light primordial black holes,''
Phys.\ Rev.\ D \textbf{84} (2011), 024028
doi:10.1103/PhysRevD.84.024028
[arXiv:1105.2303 [astro-ph.CO]].


\bibitem{Ejlli:2019bqj}
A.~Ejlli, D.~Ejlli, A.~M.~Cruise, G.~Pisano and H.~Grote,
``Upper limits on the amplitude of ultra-high-frequency gravitational waves from graviton to photon conversion,''
Eur.\ Phys.\ J.\ C \textbf{79} (2019) no.12, 1032
doi:10.1140/epjc/s10052-019-7542-5
[arXiv:1908.00232 [gr-qc]].


\bibitem{Ejlli:2018hke}
D.~Ejlli and V.~R.~Thandlam,
``Graviton-photon mixing,''
Phys.\ Rev.\ D \textbf{99} (2019) no.4, 044022
doi:10.1103/PhysRevD.99.044022
[arXiv:1807.00171 [gr-qc]].





\bibitem{Boccaletti70}
D. Boccaletti, V. De Sabbata, P. Fortini, C. Gualdi,   ``Conversion of photons into gravitons and vice versa in a static electromagnetic field ,''
Nuovo Cimento, {\bf 70B} (1970) 129.

\bibitem{Sazhin73}
 L.~P.~Grishchuk and M.~V.~Sazhin,
  ``Emission of gravitational waves by an electromagnetic cavity,''
  Zh.\ Eksp.\ Teor.\ Fiz.\  {\bf 65} (1973) 441.
  
\bibitem{Zeldovich73}
Ya.B. Zel'dovich, ``Electromagnetic and gravitational waves in a stationary magnetic field,''
Zh. Eksp. Teor. Fiz. {\bf 65} (1973) 1311 [Sov. Phys. JETP, {\bf 38} (1974) 652.

\bibitem{DeLogi77}
W.~K.~De Logi and A.~R.~Mickelson,
  ``Electrogravitational Conversion Cross-Sections in Static Electromagnetic Fields,''
  Phys.\ Rev.\ D {\bf 16} (1977) 2915.
  
\bibitem{Raffelt88}  
  G.~Raffelt and L.~Stodolsky,
  ``Mixing of the Photon with Low Mass Particles,''
  Phys.\ Rev.\ D {\bf 37} (1988) 1237. 
 

\bibitem{Dolgov:2012be}
  A.~D.~Dolgov and D.~Ejlli,
  ``Conversion of relic gravitational waves into photons in cosmological magnetic fields,''
  JCAP {\bf 1212} (2012) 003.\\
  A.~D.~Dolgov and D.~Ejlli,
  ``Resonant high energy graviton to photon conversion at the post-recombination epoch,''
  Phys.\ Rev.\ D {\bf 87} (2013) no.10,  104007.



   \bibitem{Bastianelli05}
   F.~Bastianelli and C.~Schubert,
  ``One loop photon-graviton mixing in an electromagnetic field: Part 1,''
  JHEP {\bf 0502} (2005) 069.\\
  F.~Bastianelli, U.~Nucamendi, C.~Schubert and V.~M.~Villanueva,
  ``One loop photon-graviton mixing in an electromagnetic field: Part 2,''
  JHEP {\bf 0711} (2007) 099.  











 
 





   
   
\end{thebibliography}
\end{document}